\documentstyle[psfig]{mn} 
\title{NGC~1883: a neglected intermediate-age open cluster located
in the outskirts of the Galactic disk\thanks{Based 
on observations  
carried out at Mt Ekar, Asiago, Italy. All the photometry is available 
at WEBDA database: http://obswww.unige.ch/webda/navigation.html}} 
\author[Carraro  at al.]        
{Giovanni Carraro$^1$, Gustavo Baume$^{1,2}$ and Sandro Villanova$^1$
\thanks{email: 
giovanni.carraro@unipd.it (GC); baume@pd.astro.it (GB);
villanova@pd.astro.it(SV) }\\ 
$^1$Dipartimento di Astronomia, Universit\`a di Padova, Vicolo 
dell'Osservatorio 2, I-35122 Padova, Italy \\
$^2$Facultad de Ciencias Astron\'omicas y Geof\'{\i}sicas de la UNLP,
IALP-CONICET, Paseo del Bosque s/n, La Plata, Argentina \\
 } 
 
\date{\it Submitted: March 2003} 
\pubyear{2003} 
\begin{document} 
\maketitle 
\title{The open cluster NGC~1883} 
 
\begin{abstract} 
We report on $BVI$ CCD photometry of a field centered  
in the region of the open cluster NGC~1883 down to $V=21$. 
This cluster has never been studied insofar, and we provide
for the first time estimates of its fundamental parameters,
namely radial extent, age, distance and reddening.
We find that the cluster has a radius of about 2.5 arcmin,
and shows signatures of dynamical relaxation. NGC~1883
is located in the anti-center direction, and exhibits a reddening
in the range E$(B-V)= 0.23-0.35$, depending on the metal
abundance. It turns out to be of intermediate-age 
(1 billion years old), and quite distant for an open cluster.
In fact it is located 4.8 kpc from the Sun, and more than 13 kpc
from the Galactic center. This results makes NGC~1883
one of the most peripheral old open clusters, with
important consequences for the trend of the metallicity
with distance in the outer Galactic disk.
\end{abstract} 
 

\section{Introduction} 
NGC~1883 (C~0522+465, OCL~417)  
is a northern open cluster, 
located toward 
the anti-center direction ($\alpha=05^{\rm h}~25^{\rm m}.9$,  
$\delta=+46^{\circ} 
29^{\prime}$, $l=163^{\circ}.08$, $b=+06^{\circ}.16$, J2000.).
This cluster was never studied insofar.  
The only information we have come from Trumpler (1931)
who suggests that this cluster is quite well detached
from the field, shows a clear central concentration
with a  radius of about 3 arcmin, and 
is moderately rich ({\it I2m} Trumpler class).
The same kind of information can be derived by consulting
Collinder (1931) work.\\
This paper presents the first photometric study of the cluster,
and it is part of a series dedicated at improving 
the photometry of northern open clusters at Asiago 
Observatory (see Carraro et al. 2002, and references therein)
 
\noindent
The plan of the paper is as follows. Sect.~2 illustrates  
the observation and reduction strategies. 
An analysis of  the geometrical
structure and star counts in the field of NGC~1883
are presented in Sect.~3, whereas a discussion of
the Color-Magnitude Diagram (CMD) is performed in Sect.~4.
Sect.~5 deals with the determination of cluster reddening, 
distance and age. 
Finally, Sect.~6 summarizes our findings.

\section{Observations and Data Reduction} 
 
$\hspace{0.5cm}$
CCD $BVI$ observations were carried out with the AFOSC camera at the 1.82 m 
Copernico telescope of Cima Ekar (Asiago, Italy), in the photometric night of 
November 8, 2002. AFOSC, with a pixel size of $0^{\prime\prime}.473$,  samples a $8^\prime.14\times8^\prime.14$ field in a 
$1K\times 1K$ nitrogen-cooled thinned CCD. \\

\noindent
Details of the observations are listed in Table~2 where the observed 
field is 
reported together with the exposure times, the typical seeing values and the 
air-masses. Fig.~1 shows the finding chart NGC~1883 region
taken from the DSS-2\footnote{Second generation Digitized Sky Survey,
{\tt http://cadcwww.dao.nrc.ca/cadcbin/getdss}} archive. 
The data has been reduced with the 
IRAF\footnote{IRAF is distributed by NOAO, which are operated by AURA under 
cooperative agreement with the NSF.} 
packages CCDRED, DAOPHOT, and PHOTCAL using the point spread function (PSF)
method (Stetson 1987). The calibration equations obtained by observing Landolt 
(1992) PG 0231+051 and PG 2213-006 fields observed
along the night, are:

\begin{table} 
\fontsize{8} {10pt}\selectfont
\caption{Journal of observations of NGC~1883 and standard star fields 
together with calibration coefficients (November 8, 2002).} 
\begin{tabular}{ccccccc} 
\hline 
\multicolumn{1}{c}{Field}         & 
\multicolumn{1}{c}{Filter}        & 
\multicolumn{3}{c}{Exposure time} & 
\multicolumn{1}{c}{Seeing}        &
\multicolumn{1}{c}{Airmass}       \\
 & & \multicolumn{3}{c}{[sec.]} & [$\prime\prime$] & \\ 
\hline 
 NGC 1833       & B &  600   &  60 &  5 & 2.1 & 1.015 \\ 
                & V &  300   &  30 &  3 & 2.0 & 1.018 \\ 
                & I &  300   &  30 &  3 & 2.0 & 1.020 \\
\hline
PG 0231+051     & B &  300   &  &  & 2.4 & 1.324 \\ 
                & V &   60   &  &  & 2.2 & 1.316 \\ 
                & I &   90   &  &  & 2.2 & 1.315 \\ 
\hline
PG 2213-006     & B &  150   &  & & 2.3 & 1.291 \\ 
                & V &   30   &  & & 2.3 & 1.298 \\ 
                & I &   30   &  & & 2.5 & 1.295 \\ 
\hline
\hline
Calibration     & \multicolumn {3}{l}{$b_1 = +1.602 \pm 0.004$} \\
coefficients    & \multicolumn {3}{l}{$b_2 = +0.038 \pm 0.006$} \\
                & \multicolumn {3}{l}{$b_3 = +0.29$}            \\
		& \multicolumn {3}{l}{$v_{1bv} = +1.003 \pm 0.014$} & \multicolumn {3}{l}{$i_1 = +1.691 \pm 0.044$} \\
		& \multicolumn {3}{l}{$v_{2bv} = -0.016 \pm 0.018$} & \multicolumn {3}{l}{$i_2 = +0.057 \pm 0.043$} \\
		& \multicolumn {3}{l}{$v_3 = +0.16$}                & \multicolumn {3}{l}{$i_3 = +0.08$}            \\
		& \multicolumn {3}{l}{$v_{1vi} = +1.002 \pm 0.016$} & \\
		& \multicolumn {3}{l}{$v_{2vi} = -0.013 \pm 0.016$} & \\
\hline
\end{tabular}
\end{table}

\begin{figure} 
\centerline{\psfig{file=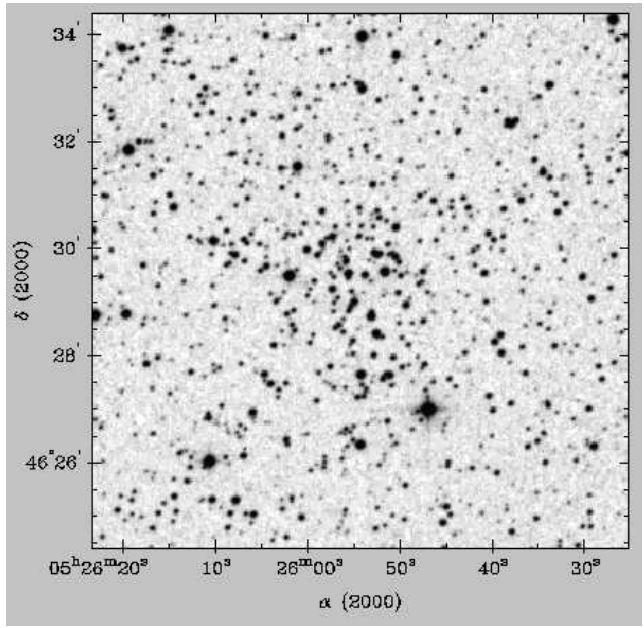,width=\columnwidth}} 
\caption{A DSS image of a region around NGC~1883
covered by the present study. North is up, east on
the left.}
\label{mappa} 
\end{figure} 
 
\begin{figure}  
\centerline{\psfig{file=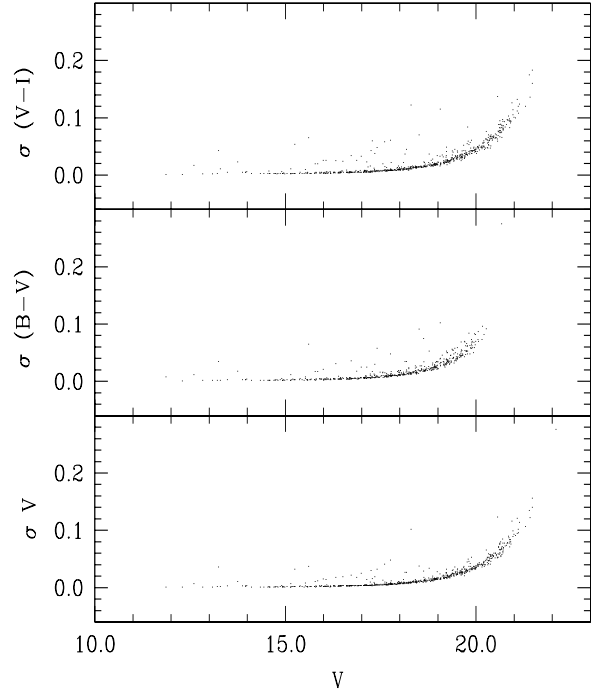,width=8cm,height=10cm}} 
\caption{Photometric errors as a function of magnitude, for our NGC~1883
observations.} 
\label{fig_errors} 
\end{figure}

\begin{center}
\begin{tabular}{lc}
$b = B + b_1 + b_2 (B-V) + b_3 X$         & (1) \\  
$v = V + v_{1bv} + v_{2bv} (B-V) + v_3 X$ & (2) \\  
$v = V + v_{1vi} + v_{2vi} (V-I) + v_3 X$ & (3) \\  
$i = I + i_1 + i_2 (V-I) + i_3 X$         & (4) \\
\end{tabular}
\end{center}

\noindent
where $BVI$ are standard magnitudes, $bvi$ are the instrumental ones, $X$ is 
the airmass and the derived coefficients are presented at the bottom of Table~2. 
As for  $V$ magnitudes, when $B$ magnitude was available, we used 
expression (2) to compute them, elsewhere expression (3) was used. The standard 
stars in these fields provide a very good color coverage.
For the extinction coefficients, we assumed the 
typical values for the Asiago Observatory (Desidera et al. 2002).

\begin{figure} 
\centerline{\psfig{file=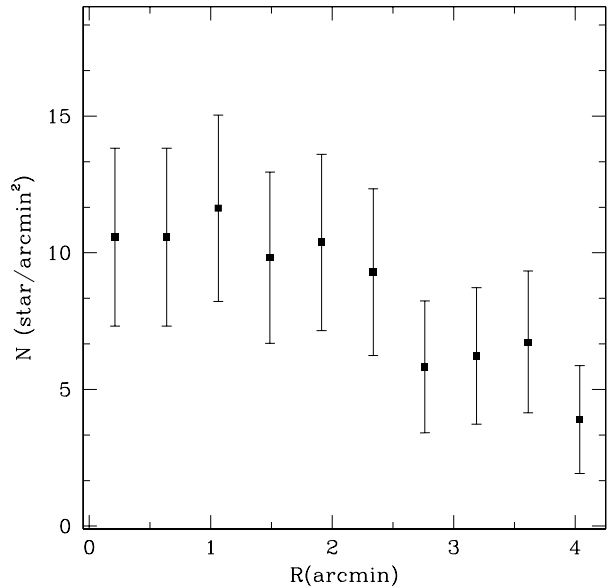,width=\columnwidth}} 
\caption{ Star counts in the field of 
of NGC~1883 as a function of the radius.}
\end{figure} 
 
\begin{figure*} 
\centerline{\psfig{file=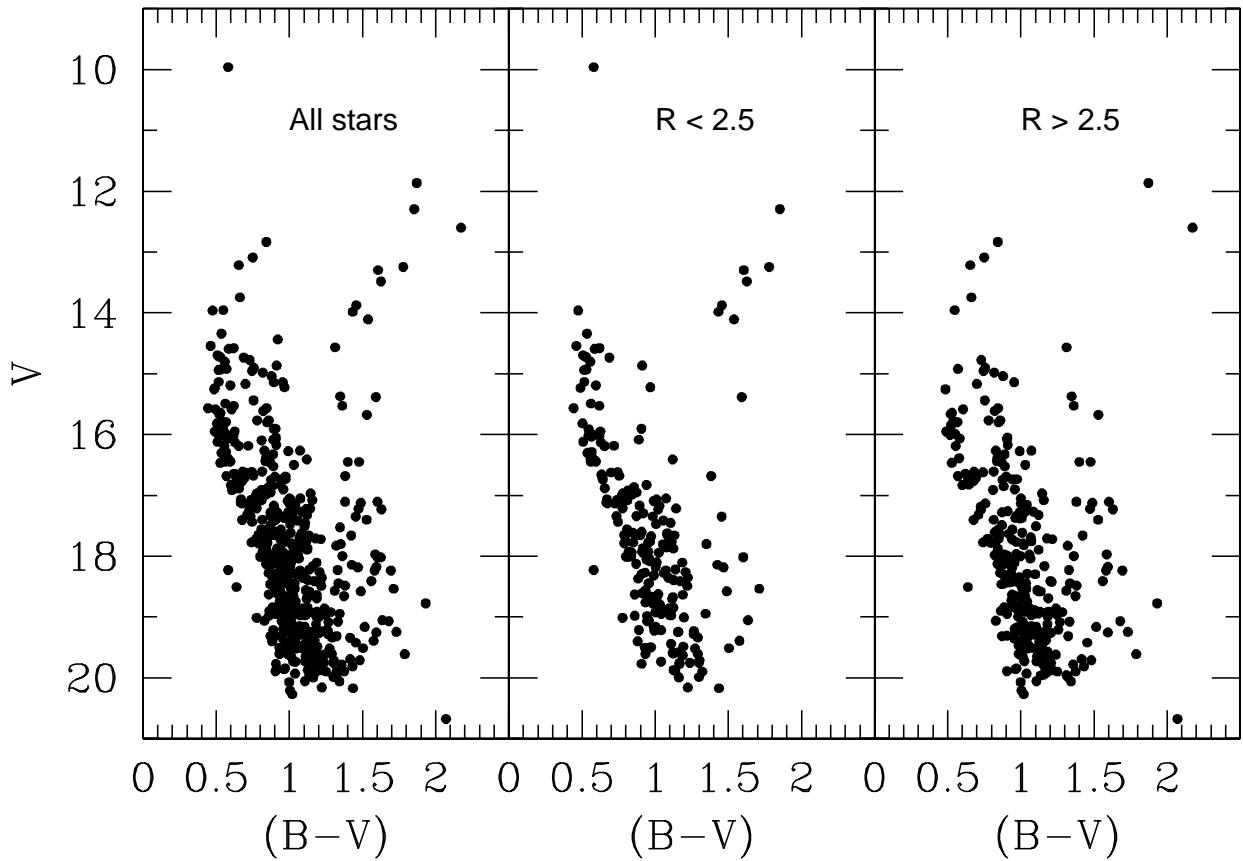,width=\textwidth}} 
\caption{CMDs of NGC~1883 as a function of the radius.} 
\end{figure*} 

Finally, Fig.~2 presents the run of photometric errors 
as a function of magnitude. These errors take into account fitting errors 
from DAOPHOT and calibration errors, and have been computed 
following Patat \& Carraro (2001). 
It can be noticed that stars brighter than  
$V \approx 21$ mag have  
photometric errors lower than 0.10~mag in magnitude
and lower than 0.15~mag in colour. The final photometric data are  
available in electronic form at the  
WEBDA\footnote{http://obswww.unige.ch/webda/navigation.html} site. \\
 
\section{Star counts and cluster size} 
Lyng\aa\ (1987) and Dias et al. (2002) report a preliminary estimate
of NGC~1883 radius, amounting to 2.5-3.0 arcmin. This is
basically confirmed by the appearance of the cluster in the finding
chart (Fig.~1), from which one can readily see that we were
able to cover all the cluster region.
We perform star counts by using our CCD data to obtain
a new estimate of the cluster size.
We derive the surface stellar density by performing star counts
in concentric rings around stars $\#12$ (the present  
numbering) selected as the approximate cluster center,
and then dividing by their
respective surfaces. The final density profile and the corresponding
Poisson error bars are shown in Fig.~3,
where we take into account all the measured stars brighter
than $V \approx 19.5$ mag.\\
The surface density  has a value of 10-12 stars/acrmin$^2$
within 2.5 arcmin, and a value of 4-6 stars/acrmin$^2$
outside this radius. 
From this sharp density variation 
we argue  that the cluster population is dominant
within 2.5 arcmin, whereas outside 
the general Galactic disk field dominates the star counts (see also Fig.~1).\\
The precise value of the cluster radius is however difficult to be obtained,
and we consider 2.5 arcmin as a lower limit. Obviously, we expect that
some cluster stars are located out of this radius, which we simply consider
as the distance from the cluster center where the cluster dominates over
the field.

\section{The Colour-Magnitude Diagrams} 
The CMDs of NGC~1883 are shown in Fig.~4.
To facilitate the interpretation we consider the distribution
of the stars in the CMD as a function of the distance from
the cluster center. In the left panel of Fig.~4 we plot all the
detected stars. Here the cluster exhibits a well
defined Main Sequence (MS) extending from $V$=20 to $V$=14.5. Interestingly,
there are a few stars in the red part of the CMD, which resemble
a Red Giant Branch (RGB). The MS is significantly wide, and
some hints for a substantial population of unresolved binary stars
is visible - particularly in the upper MS - in the form of a parallel redder
sequence. The Turn Off Point (TO) is roughly located at $V$=16,
$(B-V)$= 0.60. The group of stars above the TO ( between $V$=12.5
and $V$=13.5) can be either blue stragglers belonging to the cluster
or field stars. 
Finally the brightest star (see also Fig.~1),
located southwest of the cluster center, and recorded as GSC~03358-01188
and TYC~3358~-1188~-1, is quite probably an interloper.\\
Much better information can be obtained by looking at the middle
and right panels in Fig.~4. The middle panel contains only the stars
located inside the estimated cluster radius (2.5 arcmin), whereas
the right panel contains all the stars located outside the cluster
radius. The following considerations can be done:

\begin{itemize}
\item The MS and the TO region in the middle panel are 
much better defined;
\item almost all the probable RGB stars are inside the inner
region, which implies by the way that the cluster underwent some dynamical 
relaxation;
\item  the stars above the TO are probably field stars, since they
lie all out of the cluster radius (see right panel).
\end{itemize}

\noindent
In particular the fine shape of the TO deserves some attention. In fact
the shape of the TO is that one typical of intermediate-age open clusters,
with a blue and red hook clearly visible notwithstanding the field stars
contamination. 
This is a clear indication of an age
in the range 1-1.5 Gyr, depending on the precise metal content of the cluster
(Carraro \& Chiosi 1994, Carraro et al. 1999).

\begin{figure} 
\centerline{\psfig{file=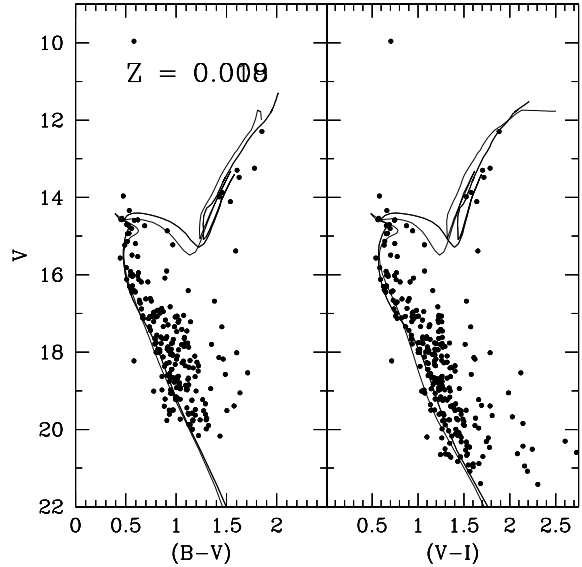,width=\columnwidth}} 
\caption{ {\bf Left panel}: NGC~1883 data in the $V$ vs.\ $B-V$ diagram,  
as compared to Girardi et al. \ (2000) isochrone of age 
$1.2\times10^9$ yr 
(solid line), for the metallicity $Z=0.019$. A distance 
modulus of $(m-M)_0=13.25$ mag, and a colour excess of E$(B-V)=0.23$ mag, 
have been derived. {\bf Right panel}: NGC~1883 data in the $V$ vs.\ $V-I$
diagram, as compared to Girardi et al. \ (2000) isochrone of age 
$1.2\times10^9$ yr 
(solid line), for the metallicity $Z=0.019$. A distance 
modulus of $(m-M)_0=13.15$ mag, and a colour excess of E$(V-I)=0.30$ mag, 
have been derived.} 
\end{figure} 
 
\begin{figure} 
\centerline{\psfig{file=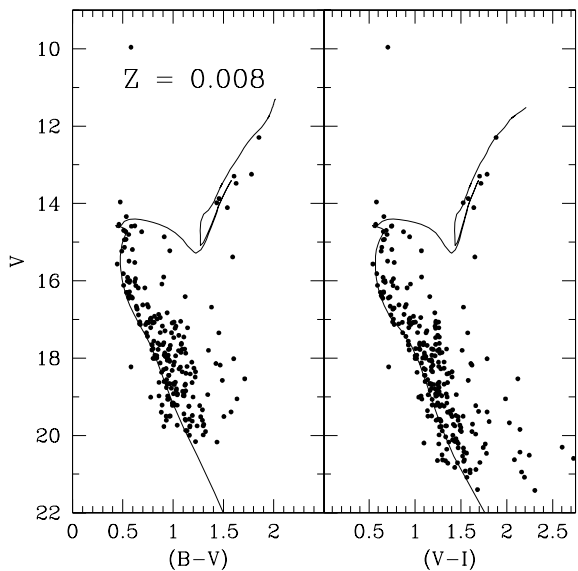,width=\columnwidth}} 
\caption{ {\bf Left panel}: NGC~1883 data in the $V$ vs.\ $B-V$ diagram,  
as compared to Girardi et al. \ (2000) isochrone of age 
$1.0\times10^9$ yr 
(solid line), for the metallicity $Z=0.008$. A distance 
modulus of $(m-M)_0=13.40$ mag, and a colour excess of E$(B-V)=0.35$ mag, 
have been derived. {\bf Right panel}: NGC~1883 data in the $V$ vs.\ $V-I$
diagram, as compared to Girardi et al. \ (2000) isochrone of age 
$1.0\times10^9$ yr 
(solid line), for the metallicity $Z=0.008$. A distance 
modulus of $(m-M)_0=13.20$ mag, and a colour excess of E$(V-I)=0.50$ mag, 
have been derived.} 
\end{figure} 

\section{Cluster fundamental parameters} 
In this section we try to provide some estimates of the cluster
basic parameters. To achieve this result, we make use of the comparison
between the stars distribution in the CMDs and a set of theoretical isochrones
from the Padova group (Girardi et al. 2000).
We already have an indication of the cluster age, but we do not know
anything about the reddening, the distance and the metallicity.
The results of the fitting are shown in Fig.~5 and 6.

\noindent
In details, in Fig.~5 we present the CMDs for the stars withing 2.5 arcmin
from the cluster center, with over-imposed an isochrone of 1.2 billion
years for a solar (Z=0.019) metallicity. The fit is quite good in the 
$V$ vs $(B-V)$ plane (left panel), but the shape of the MS is not very well
reproduced in  the $V$ vs $(V-I)$ plane (right panel). Moreover
the precise location of the RGB is not well matched.
We achieved this results by shifting the isochrone with E$(B-V)=0.23$,
E$(V-I)=0.30$, and $(m-M)=14.00$ and by adopting $R_V~=~3.1$.\\
We tried to improve the fit by lowering the metal abundance, on the base
of the fact that a cluster located toward the anti-center might have a lower
than solar metal abundance (Friel 1995). The result is shown in Fig.~6.
Here the isochrone has an age of 1.0 billion years for the Z=0.008 metal content.
The global fit is quite good, and has been obtained by shifting the isochrone
with E$(B-V)=0.35$, E$(V-I)=0.50$, and $(m-M)=14.50$ ($R_V~=~3.1$).\\
\noindent
In conclusion, a lower than solar metal abundance seems to be favored.
If this is the case, NGC~1883 turns out to be located 4.8 kpc from the Sun
toward the anti-center direction. This implies a distance from the Galactic
center somewhat larger than 13 kpc, by assuming $R_{\odot}$=8.5 kpc, and a height above 
the Galactic plane of about 500 pc. According to Friel (1995, Table~1), NGC~1883 
turns out to be one of the few intermediate-age or old open clusters 
located in the outskirts of the Galactic disk. In fact we know
only other 3 clusters (Tombaugh~2, Berkeley~20 and Berkeley~29) located
more distant than NGC~1883.
Therefore NGC~1883 might play an important role in defining the precise shape
of the radial abundance gradient in the external regions of the Galactic disk, 
which is one of the fundamental constrains of models of chemical evolution (Chiappini et al.
2001). Moreover NGC~1883 lies significantly above the Galactic plane.
The position, together with its compactness,
are probably  the main reasons for which the cluster survived until the present
time.
 
\section{Conclusions}
We have presented the first CCD $BVI$ photometric study of the 
open cluster NGC~1883. The CMDs we derive allow us to 
infer estimates of the cluster basic parameters. In detail, we find that:
 
\begin{description} 
\item $\bullet$ the age of NGC~1883 is around 1.0 Gyr; 
\item $\bullet$ the reddening $E_{B-V}$ turns out to be 0.23$\pm$0.10
by adopting a metal abundace Z=0.019, or 0.35$\pm$0.10 mag if one 
assumes Z=0.008;
\item $\bullet$ we place the cluster at about 4.8 kpc from the Sun toward 
the anti-center direction; 
\item $\bullet$ this way NGC~1883 turns out to be one of the most distant
intermediate-age open cluster. 
\end{description}

\noindent
Future work should concentrate to obtain an estimate of the cluster
metal abundance through spectra of the RGB stars. The knowledge of the
cluster metallicity, which we could not constrain very well, is of paramount
importance to better probe the trend of the metallicity in the outer
parts of the Galactic disk (Friel \& Janes 1993).

\section*{Acknowledgements} 
We acknowledges the kind night assistance by Asiago
Observatory technical staff and in particular dr. Stefano Ciroi
who secured part of the observations discussed in this work.
This study has been financed by the Italian Ministry of 
University, Scientific Research and Technology (MURST) and the Italian 
Space Agency (ASI), and made use of Simbad and WEBDA databases.

\end{document}